\documentclass[twocolumn,showpacs,showkeys,preprintnumbers,amsmath,amssymb]{revtex4}

\usepackage{graphicx}
\usepackage{dcolumn}
\usepackage{bm}

\begin{document}


\preprint{ }

\title{Warp-Drive Quantum Computation}

\author{Mikio Nakahara$^{1}$, Juha J. Vartiainen$^{2}$, Yasushi Kondo$^{1}$, 
Shogo Tanimura$^{3}$, and
Kazuya Hata$^{1}$}

\affiliation{%
$^{1}$Department of Physics, Kinki University, Higashi-Osaka 577-8502, Japan
\\
$^2$Materials Physics Laboratory, POB 2200, FIN-02015 HUT, Helsinki University of Technology, Finland
\\
$^3$Graduate School of Engineering, Osaka City University, Sumiyoshi-ku, Osaka 558-8585, Japan
\\
}%

\date{\today}

\begin{abstract}
Recently it
has been shown that time optimal quantum computation is attained using the
Cartan decomposition of a unitary matrix.
We extend this approach by noting that the 
unitary group is compact. 
This allows us to reduce the execution time of a 
quantum algorithm $U_{\rm alg}$
further by adding an extra gate $W$ to it. 
This gate $W$ sends $U_{\rm alg}$ to another algorithm $WU_{\rm alg}$
which is executable in a shorter time than $U_{\rm alg}$. We call this 
technique warp-drive.
Here we show both theoretically and experimentally that the
warp-drive reduces the
execution time of Grover's algorithm 
implemented with a
two-qubit NMR quantum computer. Warp-drive is potentially a powerful
tool in accelerating algorithms and reducing errors in any realization 
of a quantum computer.
\end{abstract}


\maketitle


Quantum computing is an emerging discipline based on encoding
information into a quantum-mechanical system \cite{ref:8, ref:9}.
There a quantum algorithm is expressed in a form of a unitary 
matrix
\begin{equation}
U_{\rm alg} = {\mathcal{T}} \exp \left[-\frac{i}{\hbar}
 \int_0^T {\mathcal{H}}(\gamma(t)) dt\right],
\end{equation}
where $\mathcal{T}$ stands for time-ordered product, $\gamma(t)$ denotes
collectively the control parameters of the Hamiltonian ${\mathcal{H}}$
at time $t$.
This unitary matrix is often implemented in terms of so-called
elementary gates, such as U(2) gates and CNOT gates \cite{ref:1, ref:10,
ref:11}.
It is possible, instead, to directly implement a given unitary matrix
without decomposing it into these elementary gates. It is expected that
this will reduce the execution time required in general.
One method for direct implementation
is to employ numerical optimization of the control parameters
of the Hamiltonian \cite{ref:12, ref:13, ref:14}.
The other method is to use the 
Cartan decomposition of the group SU($2^n$), 
to which an $n$-qubit matrix representation 
of a quantum algorithm belongs \cite{ref:2, ref:3}. 
The latter approach, which is adopted here, is successfully
demonstrated recently using an NMR quantum computer, whose pseudopure 
state is generated by cyclic permutations of state population \cite{ref:4}. 
It should be noted that exact optimal implementation of a quantum algorithm
has been achieved in holonomic quantum computation in an idealized
case \cite{ref:x}.

A warp-drive is a fictitious gadget with which two remote points in space
are connected \cite{ref:15}. It is the purpose of the present Letter
to demonstrate, both theoretically
and experimentally, that a similar 
technique may be employed to shorten the execution time of a quantum 
algorithm. 
A time optimal implementation of the quantum algorithm is equivalent
to navigating along the time optimal path from the identity operator
to the point $ U_{\rm alg} $ in SU($2^n$) by tuning the parameters
$\gamma(t)$ in the Hamiltonian. We show below that an additional 
permutation matrix $W$ of the basis vectors,
when added after $U_{\rm alg}$, sends it to a point $W U_{\rm alg}$ near
the identity matrix $I$ so that it takes a shorter time to follow the 
time optimal path connecting
$I$ with $W U_{\rm alg}$ than $I$ with $U_{\rm alg}$, see Fig.~1. 
Therefore we have ``warp-driven'' $U_{\rm alg}$ to $W U_{\rm alg}$ by 
adding $W$.
\begin{figure}[b]
\begin{center}
\includegraphics[width=4cm]{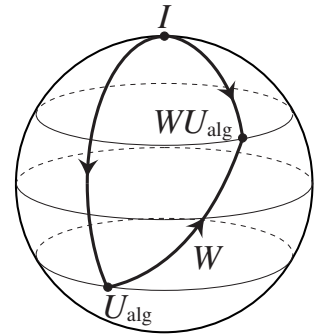}
\end{center}
\caption{Conceptual diagram showing the
effect of a warp-drive gate $W$ in a compact space SU(4).
A matrix $U_{\rm alg}$ is sent to $W U_{\rm alg}$ which is reachable from 
the unit matrix $I$ in a shorter execution time. The curves connecting
these matrices represent time optimal paths. 
}
\label{fig:sph}
\end{figure}
Although it might seem counterintuitive that the 
execution time is reduced by
adding an extra gate, this is the case since the unitary group SU($2^n$) 
is compact.
Let us consider navigating on
a sphere $S^2$, which is also a compact space. Suppose we leave from a
point $O$ on the equator toward west. Then we will eventually arrive at the
antipodal point. The distance between $O$ and us will be shorter
and shorter as we further circumnavigate the sphere. This is
what happens when an extra gate is added. 

In an NMR quantum computer, a one-qubit operation may be carried out 
in a short time on the order $ 10 \, \mu{\rm s}$ while
a two-qubit entangling operation takes time typically
$ \sim 10 \, {\rm ms}$. Thus one-qubit operation time 
may be ignored in estimating the overall execution time. 
Let us consider a molecule with two heteronucleus spins for definiteness, 
whose Hamiltonian, in the rotating frame with respective Larmor frequency, is 
\begin{eqnarray}
{\mathcal{H}}
(\gamma) &=& -\omega_{11} \left[\cos \phi_1 (\sigma_x \otimes I_2/2) + 
\sin \phi_1 (\sigma_y \otimes I_2/2) \right]\nonumber\\
& &-\omega_{12} \left[\cos \phi_2 (I_2 \otimes \sigma_x /2) + 
\sin \phi_2 (I_2 \otimes \sigma_y/2) \right]\nonumber\\
& &+ 2\pi J \sigma_z \otimes \sigma_z/4,
\label{eq:ham}
\end{eqnarray}
where $\omega_{1i}$ and $\phi_i$ are the control parameters
denoting
the amplitude and the angle in the
$xy$-plane of the external rf fields, respectively. Above $J$ denotes
the spin-spin coupling constant. Typically we have $\omega_{1i} \gg J$, 
which justifies the above assumption of negligible one-qubit
operation time compared to two-qubit operation time.
Neglecting one-qubit operation time in evaluating the execution
time amounts to 
identifying the matrices $U_1$ and $U_2$ which differ by
an element of $K \equiv {\rm SU}(2) \otimes {\rm SU}(2)$.
Thus the relevant space for
evaluating the time optimal path is the coset space
${\rm SU}(4)/{\rm SU}(2)\otimes {\rm SU}(2)$. 
To find the time optimal path connecting the unit
matrix $I$ and the matrix $U_{\rm alg}$, therefore, amounts to finding the
time optimal path connecting cosets $[I]$ and $[U_{\rm alg}]$, where $[U]
\equiv \{ k U| k \in K\}$. The Lie algebra $\mathfrak{su}(4)$ of
SU(4) is decomposed as $\mathfrak{su}(4) = \mathfrak{k} \oplus 
\mathfrak{p}$ \cite{ref:5}, where 
\begin{eqnarray}
\mathfrak{k}&=& {\rm Span}(\{i I \otimes
\sigma_j/2, i \sigma_j \otimes I/2\}),\ (j=x, y, z)\\
\mathfrak{p}&=& \mathfrak{k}^{\perp} = {\rm Span}
(\{i \sigma_j \otimes \sigma_k/4\}),\ (j,k=x, y, z).
\end{eqnarray}
Note that they satisfy the commutation relations 
\begin{equation}
[ \mathfrak{k}, \mathfrak{k}] \subset \mathfrak{k}, 
\qquad
[ \mathfrak{p}, \mathfrak{k}] \subset \mathfrak{p}, 
\qquad
[ \mathfrak{p}, \mathfrak{p}] \subset \mathfrak{k}.
\end{equation}
The decomposition of a Lie algebra $\mathfrak{g}$ into $\mathfrak{k}$ and
$\mathfrak{p}$, satisfying the above commutation relation, is called the
Cartan decomposition.
The Cartan subalgebra $\mathfrak{h} = 
{\rm Span} (\{i \sigma_j \otimes \sigma_j/4\})
\subset \mathfrak{p}$ plays an important role in the following construction.
A general theorem of Lie algebras 
proves that any element $U_{\rm alg} \in {\rm SU}(4)$ 
has a $KP$ decomposition $U_{\rm alg} = kp, k \in K \equiv \exp \mathfrak{k}$ 
and 
$p \in P \equiv \exp \mathfrak{p}$. Moreover, any matrix $p \in P$ is 
rewritten in the conjugate form $p= k_1^{\dagger} h k_1$,
where $k_1 \in K$ and $h$ is an element of the Cartan subgroup $H$
of SU(4),
\begin{equation}
H \equiv \exp \mathfrak{h} = \left\{ \exp\left(i \sum_{j=x, y, z}
\frac{\alpha_j}{4} \sigma_j \otimes \sigma_j\right)\Big| \alpha_j \in 
\mathbb{R}\right\}.
\end{equation}
Therefore we have a corresponding Cartan decomposition for a group element as
$U_{\rm alg} = k_2 h k_1$, where $k_i \in K$ and $h \in H$.
The implementation of a quantum algorithm based on a Cartan decomposition
requires shorter execution time in general \cite{ref:2, ref:3, ref:4}.

\begin{table*}[htb]
\caption{time optimal pulse sequences for Grover's algorithm $U_{10}$
and the warp-driven algorithm, $W_4 U_{10}$.
The hydrogen nucleus is the qubit 1 while
the carbon nucleus is the qubit 2.
Here X (Xm) and Y (Ym) denote $\pi/2$-pulse 
around $x$ ($-x$) and $y$ ($-y$) axis, respectively.
The symbol ${\rm Pi}(\theta)$ denotes a $\pi$-pulse around a
vector $(\cos \theta, \sin \theta, 0)$ in the Bloch sphere. 
The symbol $(1/2J)$ indicates the length of the idle time, during
which no external pulses are applied. 
The number of pulses is reduced from 10 to 4 and the execution time
is halved by adding the extra gate $W_4$.
}

\begin{tabular}{cccccccccc}
\hline
\hline
\multicolumn{1}{c}{Gate} & 
\multicolumn{8}{c}{Pulse sequence} & 
\multicolumn{1}{c}{Execution Time}
\\
\hline 
$U_{10}$  &1:  & X & ($1/2J$)& Xm &
Y  & ($1/2J$) & X & Ym &
$1/J$ \\
          &2:  & X & ($1/2J$)& Xm & 
Ym & ($1/2J$) & Y & Pi($\pi/4$)& \\
\hline 
$W_4U_{10}$  &1:  & & & Xm&
 Pi($-\pi/4$)    & ($1/2J$) & X        & &
$1/2J$ \\
             &2:  & & & & 
                 & ($1/2J$) & Pi($\pi$)& & \\
\hline
\hline
\end{tabular}
\label{table:ps}
\end{table*}

To be more concrete, let us consider implementing two-qubit Grover's 
database search algorithm $U_{ij}$ with an NMR quantum computer.
The data is encoded in one of the basis vectors $|00 \rangle,
|01 \rangle, |10 \rangle, |11 \rangle$ and the gate $U_{ij}$ picks out
a particular binary basis vector $|ij \rangle$ as a ``target file''
\cite{ref:6, ref:7}.
Here we restrict ourselves within 
$U_{10}$ which picks out the file $|10 \rangle$ with a single step.
The unitary matrix representing this algorithm takes the form
\begin{equation}
U_{10}= \left( \begin{array}{cccc}
0&1&0&0\\
0&0&0&-1\\
-1&0&0&0\\
0&0&-1&0
\end{array} \right).
\end{equation}
The Cartan decomposition of an arbitrary $U \in {\rm SU}(4)$ is carried out 
explicitly as follows. We first introduce the Bell basis \cite{ref:16}
\begin{equation}
\begin{array}{c}
{\displaystyle |\Psi_0\rangle =(1/\sqrt{2})(|00 \rangle +|11 \rangle),}
\\
{\displaystyle |\Psi_1 \rangle
= (i/\sqrt{2})(|01 \rangle +|10 \rangle),}
\\
{\displaystyle |\Psi_2 \rangle
= (1/\sqrt{2})(|01 \rangle -|10 \rangle),}
\\
{\displaystyle |\Psi_3 \rangle
= (i/\sqrt{2})(|00 \rangle -|11 \rangle).}
\end{array}
\end{equation}
The transformation rule
of a matrix $U$ with respect to the standard binary
basis $|00 \rangle,|01 \rangle,|10 \rangle,
|11 \rangle$ into that with the Bell basis $|\Psi_i \rangle$ is
$U \to U_B \equiv Q^{\dagger} U Q$, where
\begin{equation}
Q=\frac{1}{\sqrt{2}}\left(\begin{array}{cccc}
1&0&0&i\\
0&i&1&0\\
0&i&-1&0\\
1&0&0&-i
\end{array} \right).
\end{equation}
The matrix $Q$ defines
an isomorphism between $K={\rm SU}(2) \otimes {\rm SU}(2)$
and ${\rm SO}(4)$ 
and is used to classify two-qubit gates \cite{ref:16, ref:y}. 
Namely, it is easy to verify that $Q^{\dagger} k Q \in {\rm SO}(4)$
for $k \in K$.
Moreover, $Q$ diagonalizes the elements of the Cartan subgroup, viz
$Q^{\dagger} h Q = {\rm diag}(e^{i \theta_0}, e^{i \theta_1}, e^{i \theta_2}, 
e^{i \theta_3})$ for $h \in H$. Therefore we find for $U =k_2 h k_1$ that
$$
U_B = Q^{\dagger} U Q = Q^{\dagger} k_2 Q\cdot
 Q^{\dagger} h Q \cdot Q^{\dagger} k_2 Q
= O_2 h_D O_1,
$$
where $O_i \equiv Q^{\dagger} k_i Q \in {\rm SO}(4)$ and 
$h_D \equiv Q^{\dagger} h Q$ is a diagonal matrix. From $U_B^{T} U_B
= O_1^{T} h_D^2 O_1$, we notice that $U_B^{T} U_B$ is diagonalized by
$O_1$ and its eigenvalues form the diagonal elements of $h_D^2$.
Finally $O_2$ is found as $O_2 = U_B(h_D O_1)^{-1}$. 

We apply the above strategy to find the Cartan decomposition 
$U_{10}=k_2 h k_1$. An example of the time optimal control is
\begin{eqnarray}
	k_1 &=& I_2 \otimes I_2,
	\nonumber\\
	h&=& e^{i (\pi/4)(\sigma_x \otimes \sigma_x-\sigma_y \otimes \sigma_y)},
	\\
	k_2 &=& e^{-i (\pi/4) \sigma_z} \otimes e^{i(\pi/2 \sqrt{2})
	(\sigma_x + \sigma_y)}.
	\nonumber
\end{eqnarray}
To implement this decomposition with an NMR quantum computer, such terms
as $e^{i(\pi/4) (\sigma_x \otimes \sigma_x)}$ must be rewritten in favor
of the subset of generators of SU(4) in the Hamiltonian (1). We verify,
for example, that
\begin{eqnarray}
e^{i(\pi/4) (\sigma_x \otimes \sigma_x)}&=&
[e^{i(\pi/4)\sigma_x}\otimes e^{-i(\pi/4) \sigma_y}]
e^{i(\pi/4) (\sigma_z \otimes \sigma_z)}\nonumber\\
& & \times
[e^{-i(\pi/4)\sigma_x} \otimes e^{i(\pi/4)\sigma_y}].
\label{eq:trick}
\end{eqnarray}

The extra gate $W$ which possibly shortens the execution time must be
simple enough so that we can deduce the output of $U_{\rm alg}$
efficiently
from that of $W U_{\rm alg}$ using classical computation only.
We call such gates warp-drive gates. Since a matrix $U_{\rm alg}$
is an element of a compact group U($2^n$), there always exist such 
warp-drive gates which will reduce the execution time.
In the present work, we choose
the permutation matrices of the binary
basis vectors as candidates for such gates. 
Since there are four basis vectors, there are $4!=24$ permutation matrices.
Note that we are not required to examine all of 24 permutations since
18 of them are obtained by applying one-qubit rotations on the
following six permutations:
\begin{eqnarray}
W_0&=&I_4, \ W_1=U_{\rm CNOT}^{12}, \ W_2=U_{\rm CNOT}^{21},
\nonumber\\
W_3&=&U_{\rm SWAP}, \ W_4=U_{\rm cp} \equiv U_{\rm CNOT}^{12}U_{\rm CNOT}^{21},\\
W_5&=& U_{\rm cp}^2 \equiv U_{\rm CNOT}^{21}U_{\rm CNOT}^{12},\nonumber
\end{eqnarray} 
where $U_{\rm SWAP}$ represents the swap gate and
$U_{\rm CNOT}^{ij}$ is the CNOT gate whose control bit is $i$ while the
target bit is $j$. The matrices $U_{\rm cp}$ and $U_{\rm cp}^2$ have
been utilized to cyclically permute the state populations
to generate pseudopure states \cite{ref:17}.

We have optimized the execution time of the matrices $W_i U_{10}\ (0
\leq i \leq 5)$ by utilizing the Cartan decomposition outlined above
and found that
the execution time is $1/J$ for $i = 0, 1$ and $2$ while $1/2J$
for $i=3, 4$ and $5$. A time optimal control for the
warp-driven gate $W_4 U_{10}=k_2 h k_1$ is 
\begin{eqnarray}
k_1&=& e^{i (\pi/2\sqrt{2})(\sigma_x-\sigma_z)}\otimes e^{i(\pi/4) \sigma_x},
\nonumber\\ 
h &=& e^{i(\pi/4) \sigma_y \otimes \sigma_y},\label{eq:decomp}\\ 
k_2&=& I \otimes e^{i \pi \sigma_x/4}.\nonumber
\end{eqnarray}
Therefore the execution time $T$ satisfies $\pi J T/2= \pi/4$, yielding
$T = 1/2J$. 
Since the decomposition (\ref{eq:decomp})
contains generators which do not exist
in the Hamiltonian, a trick similar to (\ref{eq:trick}) must be employed.
The one-qubit gates are realized using 
rf-pulses in an NMR quantum computer.
Table I shows the actual NMR pulse sequences derived from the Cartan
decompositions of $U_{10}$ and $W_4 U_{10}$.
We call the hydrogen nucleus
and the carbon nucleus as qubit 1 and qubit 2, respectively, 
in Table I and the rest of this Letter.
In spite of the extra gate $W_4$, the latter requires less
pulses and half of the execution time required for the former.
The permutation
$W_4$ maps the binary basis vectors as $W_4: |10\rangle \mapsto |11 \rangle$, 
and accordingly the output state of
$W_4 U_{10}$ upon acting the initial state $|00 \rangle$ is $|11\rangle$.

In our experiments, a 0.6 ml, 200 mM sample of carbon-13
labeled chloroform (Cambridge Isotope) in $\mbox{d-6}$ acetone has been employed
as a two-qubit molecule and data were taken at room temperature with
a JEOL ECA-500 NMR spectrometer, whose hydrogen Larmor frequency is 
approximately 500 MHz. The measured
spin-spin coupling constant is $J = 215.5$~Hz and the transverse relaxation
time is $T_2 \sim 7.5$~s for the hydrogen nucleus and $T_2 \sim
0.30$~s for the carbon nucleus. The longitudinal relaxation time is
measured to be $T_1 \sim 20$~s for both nuclei. The initial state $|00 \rangle$
is prepared as a pseudopure state generated by the field gradient method
\cite{ref:18}.

\begin{figure}[b]
\begin{center}
\includegraphics[width=8cm]{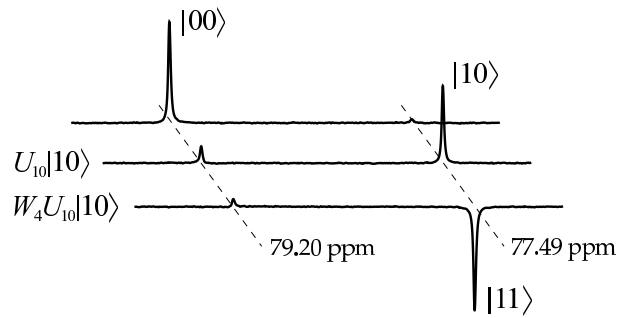}
\end{center}
\caption{
NMR spectra obtained by applying a reading pulse to the $^{13}$C
nucleus. 
The rear row shows the pseudopure initial state $|00 \rangle$
while the middle row indicates the state
$|10\rangle$ obtained by executing $U_{10}$ on $|00\rangle$. 
Finally the front row shows the spectrum obtained from warp-drive 
quantum computing $W_4 U_{10}$, where the peak corresponds to
$|11 \rangle$.
}
\label{fig:s}
\end{figure}

Figure 2 shows our experimental result. The spectra for the $^{13}$C 
nucleus (qubit 2) are shown in the panels. The spin states of both
nuclei are inferred from these spectra.
The position of the peak depends on the state of the qubit 1: 
peaks at 77.49 ppm (part per million) and 79.20 ppm 
stand for the states $|1 \rangle$ and $|0\rangle$, 
respectively, of the qubit 1.
The positive (negative) amplitude indicates qubit 2 being 
in the state $|0\rangle$ ($|1\rangle$). 
Thus the peak in the rear row shows the initial pseudopure state
is $|00 \rangle$ while the peak in the middle row tells us
that the execution of Grover's algorithm $U_{10}$ generates the 
state $|10 \rangle$. 
Finally the front row
shows the spectrum obtained after the execution of the
warp-driven gate $W_4 U_{10}$.
The peak shows that the resulting state is $|11 \rangle$, from which
the output of $U_{10}|00 \rangle$ is easily deduced as $|10\rangle$.
The spectrum obtained with the warp-driven
gate is sharper than that with the original $U_{10}$ gate.
The undesirable peak at $|00 \rangle$ is also improved by the warp-drive.
Smaller number of gates and reduced execution time
account for these improvements.

%
We have also examined the effect of the warp-drives for
Grover's algorithms other than $U_{10}$
and found that the execution times remain unchanged for $W_{0,1,2}$
while they are halved for $W_{3,4,5}$. It should not be expected, however,
that
the latter gates reduce the execution time of an arbitrary two-qubit gate:
clearly, such a ``universal warp-drive'' which reduces execution time
of an arbitrary quantum algorithm does not exist.
In view of the compactness of the unitary group, however, there exists
a finite set of simple unitary transformations, such that the execution time
of any quantum algorithm is reduced by properly chosen elements of the set.
We expect that the set of permutation matrices, or certain subset thereof,
is such a ``universal warp-drive set''.

In summary, we have demonstrated
both theoretically and experimentally that
a quantum algorithm may be accelerated by adding an extra gate to it. 
We took advantage of the compactness of the group SU(4) and found the
time optimal control parameters utilizing the Cartan decomposition for it.
When applied to two-qubit
Grover's algorithm in the NMR experiment the execution time
is found to be halved along with a reduction in the number of the
gate pulses from
10 to 4. As a result, application of warp-drive quantum computing sharpens
the NMR spectrum and reduces the spurious peak,
implying reduction in decoherence and gate operation
errors. How reductions in the execution time and the number of gate pulses
scales with the number of qubits remains a challenging problem.


MN would like to thank partial supports of Grant-in-Aids for 
Scientific Research from the Ministry of Education, 
Culture, Sports, Science and Technology, Japan, Grant No.~13135215
and
from Japan Society for the Promotion of Science (JSPS), Grant No.~14540346.
JJV would like to thank Nokia Foundation for support. 
ST is partially supported by JSPS, Grant No.~15540277.

\end{document}